\documentclass[%
reprint,
amsmath,amssymb,
aps,
prl,
]{revtex4-2}

\usepackage{graphicx} 
\usepackage{dcolumn} 
\usepackage{bm} 
\usepackage{hyperref} 
\usepackage{color}
\usepackage{soul}
\usepackage{float}
\usepackage[dvipsnames]{xcolor}
\usepackage{physics}
\usepackage{ulem}
\graphicspath{{./}}

\graphicspath{ {./} }

\begin{document}

\title{Kerr--optomechanical spectroscopy of multimode diamond resonators}

\author{Parisa Behjat,  Peyman Parsa, Natalia C. Carvalho, Prasoon K.\ Shandilya, and Paul E.\ Barclay$^*$}

\address{Institute for Quantum Science and Technology, University of Calgary, Calgary, Alberta T2N 1N4, Canada}
\email[Paul~E.\ Barclay: ]{Corresponding author pbarclay@ucalgary.ca}
\date{\today}

\begin{abstract}
Diamond microdisk cavities play a key role in optomechanical and spin-optomechanical technologies. Previous optomechanical studies of these devices have focused exclusively on their fundamental radial breathing mode. Accessing other mechanical modes of these structures is desirable for identifying routes towards improving their optomechanical properties, implementing multimode optomechanical systems, and broadening the accessible range of resonant spin--phonon coupling processes. Here we perform broadband optomechanical spectroscopy on diamond microdisks, and observe high quality factor mechanical modes with frequencies up to 10 GHz. Through Fano interference of their optomechanical response with diamond's Kerr nonlinear optical response, we estimate that optomechanical coupling of these high frequency modes can exceed 10 kHz, making them attractive for high-frequency multimode optomechanics. In combination with their per-phonon stress of a few kPa, these properties makes them excellent candidates for spin-optomechanical coupling.
\end{abstract}

\maketitle


\section{Introduction}

Nanoscale optical resonators enhance the coupling between light and matter by confining photons to wavelength scale volumes, creating large per-photon fields. They underpin breakthroughs in nonlinear optics \cite{kippenberg2018dissipative}, solid-state and atomic quantum optics \cite{wolfowicz2021quantum, chang2014quantum}, sensing \cite{yu2021whispering}, and quantum optomechanics \cite{aspelmeyer2014cavity, barzanjeh2022optomechanics}. Optical resonators created from diamond have attracted considerable attention, in part thanks to diamond's ability to host defects \cite{ref:jelezko2006sdc} whose electronic and nuclear degrees of freedom are used as memories in quantum networks \cite{pfaff2014unconditional, stas2022robust} and as quantum sensors \cite{ref:balasubramanian2008nim, ref:maze2008nms}. Diamond's optical and thermal properties allow it to support intense optical fields, leading to enhancement of nonlinear optical effects fueling demonstrations of microresonator based optical frequency combs \cite{hausmann2014diamond, lux2014multi} and lasers \cite{williams2018high}. These properties, in combination with diamond's exceptional mechanical characteristics, make it an ideal material for creating cavity optomechanical devices that coherently couple optical and mechanical resonances \cite{shandilya2022diamond}.

Cavity optomechanical devices have led to breakthroughs in quantum phononics \cite{barzanjeh2022optomechanics} including demonstration of quantum memories \cite{wallucks2020quantum} and transducers \cite{brubaker2022optomechanical, mirhosseini2020superconducting, jiang2020efficient}. Diamond cavity optomechanical devices \cite{burek2016diamond, mitchell2016single} have broadened the toolbox available for quantum and classical information processing by introducing coupling of photons and phonons to quantum spin systems associated with defects in the diamond crystal \cite{shandilya2021optomechanical}. Optomechanical functionality of diamond microdisks has been expanded by harnessing more than one of their regularly spaced optical resonances, which span visible to IR wavelengths, and whose shared optomechanical coupling to a single mechanical resonance of the device has enabled wavelength conversion \cite{mitchell2019optomechanically}, optical switching \cite{lake2020two}, and amplification enhanced optomechanical memory \cite{lake2021processing}.

Incorporating coupling to multiple mechanical modes promises to further broaden the range of applications of diamond microdisks. Coupling to multiple mechanical modes underlies recent demonstrations of topological states \cite{habraken2012continuous},  Floquet phonon lasing \cite{mercade2021floquet}, multimode interference \cite{kuzyk2017controlling} and state transfer \cite{weaver2017coherent}, and entanglement of two mechanical resonators \cite{kotler2021direct}. Previous studies of diamond microdisks have been limited to its mechanical radial breathing mode. This is in part due to this mode's strong optomechanical interaction with microdisk whispering gallery modes. However, in addition to enabling multimode applications, other microdisk mechanical modes may have stronger mechanical strain concentration, an important figure of merit for spin-optomechanics \cite{shandilya2021optomechanical},  as well as higher mechanical frequency and lower mechanical dissipation, which can lead to improved resilience to thermal decoherence \cite{aspelmeyer2014cavity, safavi2019controlling}.

In this work we experimentally observe and characterize a broad spectrum of diamond microdisk mechanical modes for the first time. Using drive and probe lasers coupled to separate optical modes, we resonantly excite the motion of mechanical resonances over a bandwidth spanning 1--10 GHz. In addition to measuring their mechanical susceptibility and corresponding mechanical dissipation, we assess their optomechanical coupling strength. This latter measurement is aided by coherent Fano interference between resonant optomechanical and broadband Kerr nonlinear optical modulation of the cavity mode dynamics, which allows the optomechanical coupling strength to be measured independently of the system's optical response.

\section{Optomechanical spectroscopy}

\begin{figure}[tb]
	\includegraphics[width= 1.0\linewidth]{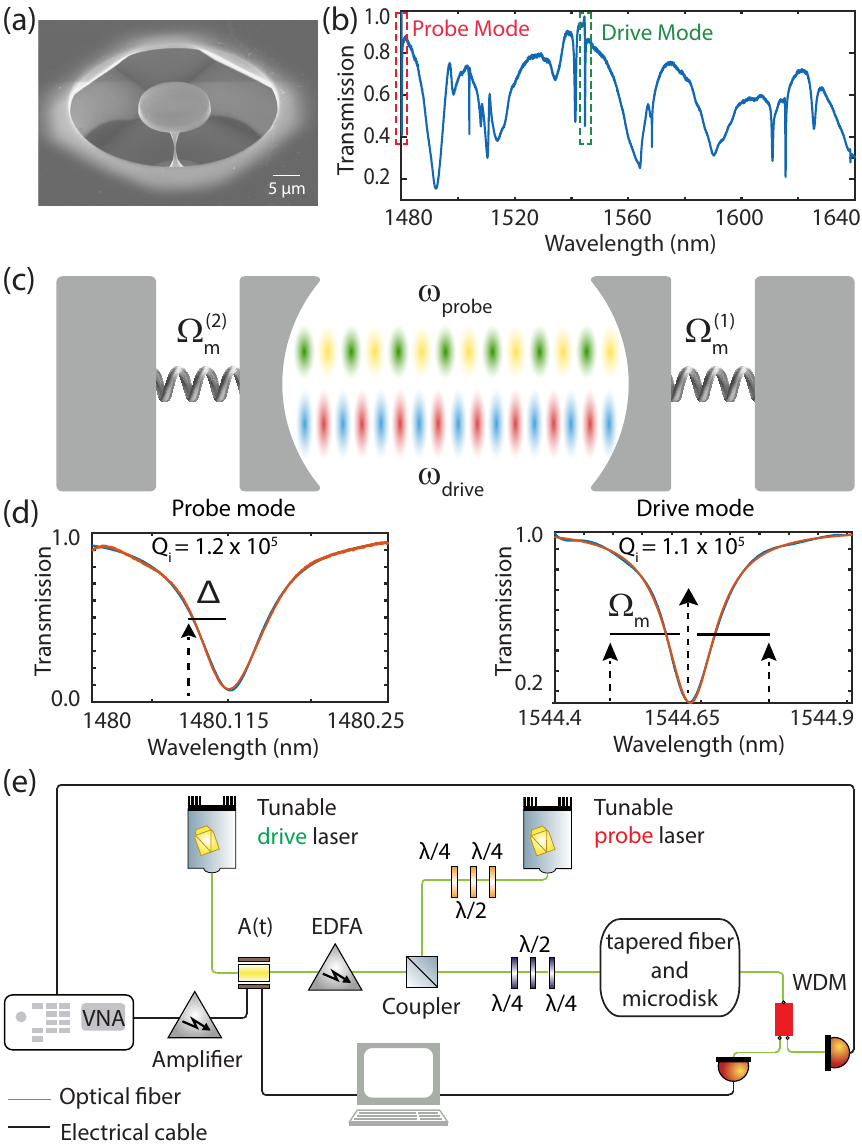}
		\caption{(a) SEM image of a diamond microdisk studied here. (b) Normalized fiber taper transmission as a function of input laser wavelength. Red and green dashed lines highlight the optical modes used for the probe and drive fields, respectively. (c) Schematic of the multimode cavity optomechanical system studied using the two mode measurement scheme.  Amplitude modulation of the drive field excites mechanical modes of the cavity. A weak probe field transduces motion of the mechanical resonatonces of the cavity. (d) Transmission spectra (red) and Lorentzian fits (blue) of the probe and drive modes. (e) Experimental apparatus. WDM: wavelength division multiplexer used to separate the drive and probe fields. The amplitude modulation follows $A(t)$.}
		\label{f1}    
\end{figure}

The diamond microdisk cavity optomechanical system studied here is shown in the SEM image in Fig.\ \ref{f1}(a) and is similar to those studied in previous work \cite{mitchell2016single}. It supports optical whispering gallery modes that couple through radiation pressure and the photoelastic effect to mechanical resonances of the microdisk structure. Previous studies focused exclusively on optical coupling to the radial breathing mode, whose mechanical displacement modulates the microdisk diameter and corresponding optical path length of its whispering gallery modes. In this work, we use two optical modes in a pump--probe setup to drive and observe motion of other microdisk mechanical resonances. The device's optical mode spectrum, measured from the transmission of a tunable laser input to a fiber taper waveguide evanescently coupled to the device, is shown in Fig.\,\ref{f1}(b). As illustrated by the canonical cavity-optomechanical system in Fig.\ \ref{f1}(c), each of these modes interacts with multiple mechanical resonances of the cavity. The drive and probe modes are highlighted in Fig.\ \ref{f1}(b) and high-resolution scans of their lineshapes are shown in Fig.\,\ref{f1}(d).

The cavity's optomechanical response was probed using the experimental apparatus shown in Fig.\,\ref{f1}(e).  A continuous wave laser input to the drive mode is modulated by an electro-optic amplitude modulator (EOSpace AZ-0K5-10-PFA-SFA) driven by the RF output of a vector network analyzer (VNA: Keysight E5063A) and then amplified with an erbium doped fiber amplifier (EDFA: Pritel LNHP-FA-20-IO-CP). The resulting oscillating intracavity field excites mechanical modes of the microdisk. Simultaneously, the transmission of a weak laser input to the probe mode is measured. This field transduces mechanical motion of the microdisk as well as other modifications to the cavity dynamics induced by the drive field. 

Both the drive and probe modes are strongly coupled to the fiber taper waveguide.  Together with parasitic losses induced by the fiber taper \cite{spillane2003ideality}, this coupling reduces the drive mode's quality factor from its intrinsic value $Q_\text{i}^\text{d}= 1.1 \times 10^{5}$ to a loaded value of $Q_\text{o}^\text{d}= 1.5 \times 10^{4}$. Similarly, the probe mode is characterised by $Q_\text{i}^\text{p}= 1.2 \times 10^{5}$ and $Q_\text{o}^\text{p}= 2.5 \times 10^{4}$. Operating in this regime increases the cavity optical bandwidth to $\kappa_{\text{p,d}} = \omega_\text{p,d}/Q_\text{o}^{\text{p,d}} \sim 2\pi \times 10\,\text{GHz}$, where $\omega_\text{p,d}$ are the probe and drive mode frequencies, respectively. This wide bandwidth is beneficial to optomechanically driving high-frequency mechanical resonances.

\begin{figure}[tb]
	\includegraphics[width=1.0\linewidth]{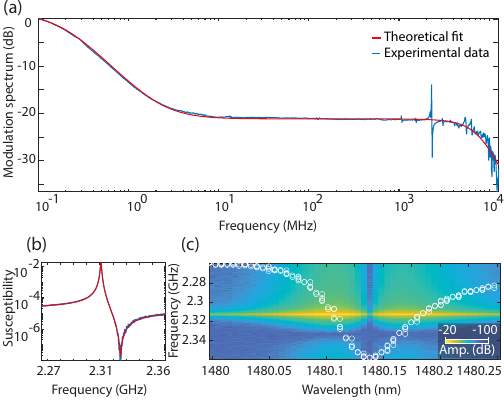}
	\caption{(a) Photodetected probe field spectrum as a function of drive field modulation frequency. The probe modulation is dominated by the photothermal effect for $\Omega/2\pi < 10\,\text{MHz}$. Beyond this point the optical Kerr effect dominates, giving rise to the middle plateau. Above a few GHz the optical mode bandwidth limits the transduced response, leading to the second drop.  The spikes in the experimental data (blue) belong to the optomechanical response of various mechanical modes. (b) Driven response (susceptibility) of the first order radial breathing mode (2.31 GHz) with the probe detuning set to maximize the transduction amplitude. (c) Dependence of the susceptibility of the radial breathing mode on probe laser detuning.}
		\label{f2}	  
\end{figure}

The cavity spectrum obtained when the drive frequency is swept from $\Omega{/2\pi} \sim 100\,\text{kHz} - 12\,\text{GHz}$ is shown in Fig.\  \ref{f2}(a). In this measurement, the drive laser is on-resonance to maximize its intracavity field, and the probe laser is detuned from resonance by  $\Delta_\text{p}/2\pi = \kappa_\text{p}/(2\sqrt{3})$ to maximize transduction of the dominantly dispersive coupling between the mechanical resonator and the probe mode resonance frequency.  The spectrum exhibits a broad background with three distinct regions. For $\Omega/2\pi < 10\,\text{MHz}$ the signal is dominated by the device's photothermal response and falls with increasing frequency due to the system's finite thermal time constant. For $\Omega/2\pi > 3\,\text{GHz}$ the signal falls with increasing frequency due to the finite optical bandwidth of the drive and probe cavity modes. Between these two regions the signal background is at a constant value related to the cavity's nonlinear optical response \cite{rokhsari2005observation, lu2014optical, lu2015high, gao2022probing}, as described below. Superimposed upon this background are contributions from the microdisk's mechanical resonances. The highest amplitude resonance is near $\Omega/2\pi = 2.3\,\text{GHz}$ and corresponds to motion of the radial breathing mode studied in previous work \cite{mitchell2016single}. The Fano nature of its lineshape is highlighted in Fig.\ \ref{f2}(b), and its dependence on probe wavelength, shown by the spectrograph in Fig.\ \ref{f2}(c), follows the slope of the optical cavity mode lineshape, as expected for dispersive optomechanical coupling.

The Fano lineshapes transduced by the probe can be explained by interference between modulation of the cavity's optical mode frequency through changes in refractive index and changes in mechanical displacement or strain \cite{lu2015high}. This interplay is captured by an effective mechanical susceptibility of the resonator, 
\begin{equation}\label{eqn:4}
    \chi_\text{m}^\text{eff}(\Omega)=\chi_\text{m}(\Omega)-\frac{\gamma \rho}{2g_0^\text{p}g_0^\text{d}\Omega_\text{m}}.
\end{equation}
%
The first term in Eq.\ \ref{eqn:4} represents the Lorentzian mechanical susceptibility of the microdisk resonance of interest, $\chi_\text{m}(\Omega) =(\Omega_\text{m}^2-\Omega^2-i\Gamma_\text{m}\Omega)^{-1}$, where $\Omega_\text{m} = 2 \pi f_\text{m}$ is the mechanical resonance frequency and $\Gamma_\text{m}$ is the mode's mechanical dissipation rate. The second term is proportional to the instantaneous shift of the probe mode frequency induced by the oscillating drive field's modification of the cavity's refractive index via the Kerr effect. This latter term is normalized by the vacuum optomechanical coupling rates $g_0^\text{p}$ and $g_0^\text{d}$ of the probe and drive modes, respectively. As described in the Supplementary Information, $\rho = \frac{q\hbar\omega}{V_\text{eff}}$ is the single-photon frequency shift of the drive mode due to the Kerr effect, and $\gamma$ is a cross-phase modulation factor between 0 and 2, depending on the spatial overlap of the probe and drive modes.  $V_\text{eff}$ is the effective mode volume of the drive mode. The material's nonlinear optical properties determine $q = \frac{c\omega n_2}{n_0 n_g}$, where $n_2$ is the second-order nonlinear refractive index coefficient of diamond, $n_0$ is the refractive index of diamond, and $n_g$ is the group index of the drive mode.  

Using the effective mechanical susceptibility, we can describe the signal transduced by the VNA, which measures the sibeband amplitude of the intracavity probe field: 
\begin{multline}
    \label{eqn:ap}
a_\text{p}^{\pm}(\Omega) \propto i \bar{a}_\text{p} \chi_\text{o}^{\text{d}}(\omega_\text{d} \pm \Omega)\chi_\text{o}^{\text{p}}(\omega_\text{p} + \Delta_\text{p} \pm \Omega)\times \\ \left(2g_0^\text{p}g_0^\text{d}\Omega_\text{m}\chi_\text{m}^\text{eff}(\pm \Omega) + \chi_\text{T}(\pm \Omega) \right).  
\end{multline}
Here $\bar{a}_\text{p} = (\sqrt{\kappa_\text{ex,p}}\alpha_\text{in,p})/(\kappa_\text{p}/2-i\Delta_\text{p})$ is the steady state amplitude of the intracavity probe field, and $\chi^\text{d,p}_\text{o}(\omega) = \left(\kappa_\text{d,p}/2-i(\omega - \omega_\text{d,p}) \right)^{-1}$ are the optical susceptibilities of the drive and probe modes, whose finite bandwidth reduces the sideband amplitudes when $\Omega > \kappa$. We have assumed that the drive field is on-resonance ($\Delta_\text{d} = 0$).  Sideband generation from photothermal dispersion induced by optical absorption of drive photons is described by $\chi_\text{T}(\Omega) = -\alpha \Omega_\text{T}/(-i\Omega + \Omega_\text{T})$, where $\Omega_\text{T}$ is the microdisk's thermal bandwidth and $\alpha$ is the photothermal shift of the optical mode frequency per photon and is related to the cavity's optical absorption rate as well as its thermal properties \cite{gao2022probing}. This model neglects dynamic backaction from both the drive and probe fields, which is small for the sideband unresolved system used here. Photothermal driving of the mechanical modes is also not considered; such an effect can effectively enhance the optomechanical coupling rate but is negligible for $\Omega_\text{m} \gg \Omega_{T}$. The measured VNA signal amplitude is $|S_\text{21}| \propto \left|\bar{t}_\text{p}^* a_\text{p}^{+} + \bar{t}_\text{p} (a_\text{p}^{-})^*\right|$, where $\bar{t}_\text{p} = \alpha_\text{in,p}(\kappa_\text{p}/2-\kappa_\text{ex,p} - i\Delta_\text{p})/(\kappa_\text{p}/2-i\Delta_\text{p})$ is the steady state amplitude transmission coefficient of the probe field through the fiber taper.

By fitting the broadband frequency response in Fig.\ \ref{f2}(a) with this model, without considering any mechanical resonances, we see good agreement over the full modulation range. Note that the contribution from $\chi_\text{T}$ is not significant for $\Omega/2\pi > 10\,\text{MHz} \sim \Omega_\text{T}/2\pi$, consistent with the discussion above and previous measurements of the thermal time constant of these devices \cite{lake2018optomechanically}.  The broadband spectrum is normalized by a background that accounts for the frequency response of the electro-optic amplitude modulator and the photodetector.

Using Eq.\ \eqref{eqn:4}, we see that given the optomechanical coupling rates for a  combination of mechanical resonance and optical drive and probe modes, $n_2$ can be extracted from the response of the probe field, and vice versa \cite{lu2015high}. Conveniently, this measurement is independent of the drive and probe mode dynamics, as well as optical power level and other apparatus specific factors. This process is simplified for mechanical resonances falling into the frequency range above the thermal frequency cutoff, as $\left|\chi_\text{T}\right| \propto 1/\Omega$ and becomes small compared to the mechanical and nonlinear optical response.  Note that related interference effects have been observed in piezoelectronically actuated aluminum nitride microdisks \cite{han201510}.

We validate this model by fitting the response shown in Fig.\ \ref{f2}(b) of the radial breathing mode, whose optomechanical coupling rate has been studied previously \cite{mitchell2016single, lake2018optomechanically, lake2020two, lake2021processing}. We then use Eq.\ \eqref{eqn:4} to estimate $\sqrt{g_0^\text{p} g_0^\text{d}}$ of the newly observed mechanical resonances. Using the frequency noise calibration technique \cite{gorodetksy2010determination}, we measure a vacuum optomechanical coupling strength $g_0^\text{d}/2\pi=25$ kHz between the first order mechanical radial breathing mode and the drive mode, consistent with previously measured values \cite{mitchell2016single}. Assuming that the probe and drive modes have the same $g_0$ since they are part of the same mode family and only differ by azimuthal mode number (see Supplementary Information), from a fit of the Fano lineshape in Fig.\ \ref{f2}(a) with Eq.\ \eqref{eqn:4}, we extract a Kerr coefficient $n_\text{2}= 2.41 \times 10^{-20} \text{m}^2/\text{W}$. This value is in good agreement with previously reported results \cite{hausmann2014diamond}, indicating that given $n_2$, fits to the Fano response can be used to calibrate optomechanical coupling.

\begin{figure*}[tb]
	\includegraphics[width=1\linewidth]{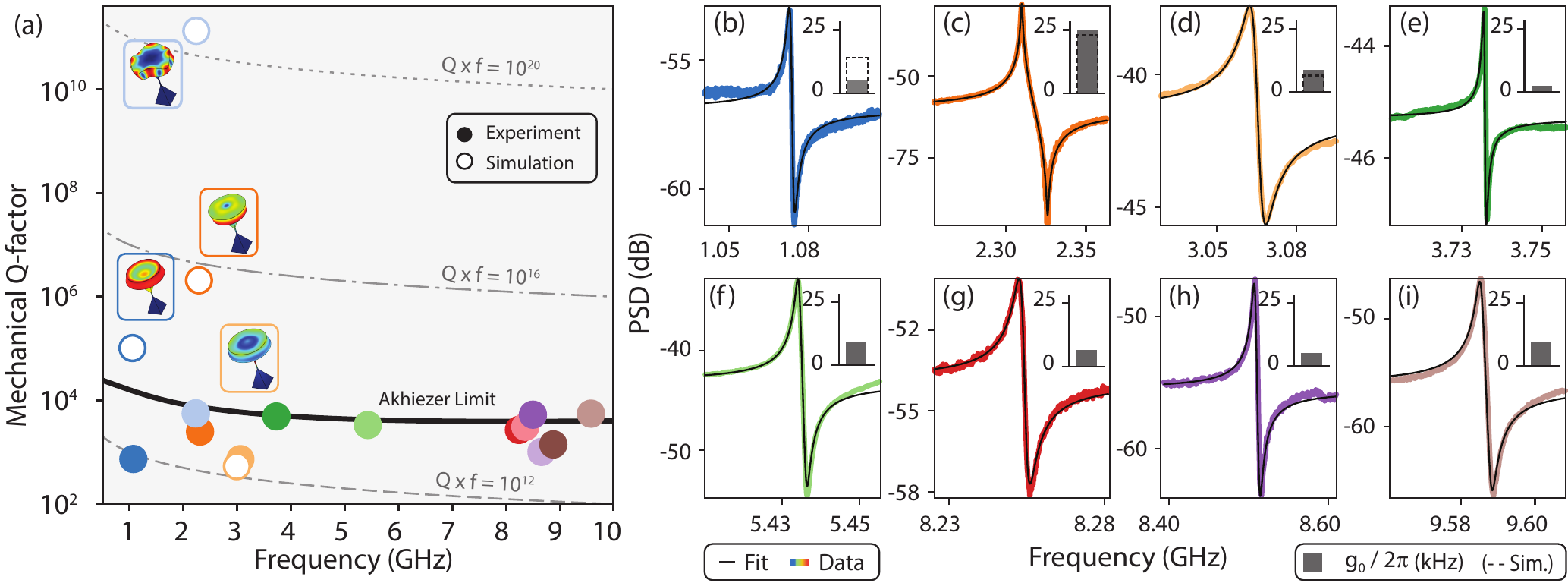}
		\caption{(a) Frequencies and quality factors of mechanical modes  measured experimentally (solid circles) and predicted from simulations (empty circles). When simulated and experimental circles match in color they correspond to the same mechanical resonances. The experimental and simulated mechanical $Q_\text{m}$ were calculated using Eq.\ \eqref{eqn:4} and a three dimensional finite element model employing perfectly matched layers, respectively. Insets show the mechanical mode profiles generated simulations for modes that were matched with experimental measurements (red and blue represent maximum and minimum displacement, respectively). Gray dashed lines show contours of constant $Q_\text{m} \times f_\text{m}$  product and the thick black line shows $Q_\text{m}$ predicted by the Akhiezer damping model at room temperature. (b) - (i) The measured susceptibility of some of the observed Fano resonances. The line colors correspond to the data presented in (a), and the black line shows a fit using model presented in the text. The insets show the optomechanical coupling rate, $g_0/2\pi$, in kHz. The dark gray bars represent the measured $g_0$, obtained by fitting the data with Eq.\ \ref{eqn:4}. For the modes in (b)--(d) that could be matched unambiguously with simulations, the dashed empty bars show the calculated $g_0$. } 
		\label{f3}   
\end{figure*}

\section{Probing higher order modes}

The additional Fano resonances in Fig.\ \ref{f2}(a) between $1 - 10$ GHz were not observed in previous measurements of diamond microdisks reliant upon thermomechanical noise spectroscopy. To assess whether these new mechanical modes are suitable for experiments in coherent and spin-optomechanics, we use the model described above to extract their optomechanical coupling strength and $Q_\text{m}$. Figure \ref{f3}(a) plots $Q_\text{m}$ obtained from fits to these resonances using Eq.\ \eqref{eqn:4}, some of which are shown in Figs.\ \ref{f3}(b) -- (i). When possible, data points are compared with $Q_\text{m}$ predicted from finite element method (FEM) simulations (performed using COMSOL), which account for radiation losses and anchoring of the microdisk to the pedestal \cite{mitchell2016single}. For frequencies above 3 GHz, a large number of modes with $Q_\text{m}$ near the thermal limit discussed below are observed. In this range, simulation and data present a high density of modes that cannot be unambiguously identified, preventing comparison with their clamping loss limited $Q_\text{m}$.
Also shown in Figs.\ \ref{f3}(b)--(i) are the extracted optomechanical coupling rates $g_0$ for each mode obtained from fitting the Fano resonance with Eq.\ \ref{eqn:4} using the value of $n_2$ obtained in the analysis of the fundamental radial breathing mode. When possible (Figs.\ \ref{f3}(b)--(d)), we compare these $g_0$ values with FEM calculated values obtained from the overlap of the optical and mechanical mode profiles, as described in \cite{primo2020}. We find good agreement for the 2.3 GHz fundamental radial breathing mode, and observe several unidentified high frequency modes with smaller but significant $g_0/2\pi \sim 10$ kHz. We have again assumed that $g_0$ is the same for both the probe and drive modes. 

In the room temperature environment used for the measurements reported here, high frequency modes can also be significantly damped by phonon--phonon interactions \cite{c2021}. Using Akhiezer phonon-scattering theory we modeled the strength of this damping mechanism (details in the Supplementary Information), which is shown as a function of frequency in Fig.\ \ref{f3}(a).  Comparing the calculated theoretical contribution to $Q_\text{m}$ from geometric effects, represented by the empty circles in Fig.\  \ref{f3}(a), with the predicted Akhiezer damping, we see that the three lowest frequency modes, including the first-order radial breathing mode (dark orange), are expected to be limited by phonon--phonon damping. In contrast, the fourth mode (light orange) is expected to have $Q_\text{m}$ limited by the anchoring losses. This comparison indicates that low-temperature operation may significantly enhance $Q_\text{m}$ for the modes limited by thermal dissipation: as the thermal phonon population is reduced, phonon--phonon scattering is suppressed and $Q_\text{m}$ will approach its geometric limit (see Supplementary Information). For example, $Q_\text{m}$ of the first-order radial breathing mode can be increased by two orders of magnitude according to Fig.\ \ref{f3}(a). This result, however, conflicts with previous studies of similar devices \cite{mitchell2016single}, where measurements of this mode's $Q_\text{m}$  for microdisks with varying pedestal width indicated that it was limited by clamping losses. Similarly, $Q_\text{m}$ of the 1 GHz mode is expected to have an Akhiezer limited value that is higher than observed in the measurements. An explanation for this discrepancy may be the pedestal's relatively large contribution to this mode's effective mass, which may in turn cause this mode to be more sensitive to discrepancies between the simulated and actual pedestal dimensions and material parameters. For example, in the Supplementary Information, we show that predictions of $Q_\text{m}$ can be impacted by diamond's Poisson ratio, which in turn varies with temperature and material properties such as defect concentration and internal stress.  The diamond material properties used in our numerical and theoretical calculations can be found in Refs.\ \cite{stoupin2010, reeber1996, migliori2008, sorokin2015}. Given the sensitivity of $Q_\text{m}$ to geometry and material properties, further optimization of pedestal shape \cite{lake2018optomechanically} has the potential to significantly increase $Q_\text{m}$, particularly when operating at low temperature where Ahkiezer damping is reduced.


\section{Discussion}
Modes with high $g_0$ and high $Q_\text{m}$ are desirable for achieving large optomechanical cooperativity $C_\text{om} = 4 N g_0^2/\kappa\Gamma_\text{m}$, where $N$ is the number of photons circulating inside the cavity. Similarly, modes with large $Q_\text{m} \cdot f_\text{m}$ are more resilient to thermal decoherence and can more easily satisfy the requirement for coherently generating a single phonon faster than the resonator's thermal decoherence, $C_\text{om} > n_\text{th}$, where $n_\text{th}$ is the thermal occupancy of the mechanical mode at a given operating temperature. High $\Omega_\text{m}$ has the additional benefit of aiding in reaching the sideband resolved regime $\Omega_\text{m} > \kappa_\text{o}$, which is a necessary condition for implementing many schemes in quantum optomechanics such as cooling and coherent phonon generation \cite{aspelmeyer2014cavity, safavi2019controlling}.  Among the measured high-frequency modes, those shown in Fig.\ \ref{f3}(h) and (i) have significant potential for future study, as they respectively combine high measured $g_0/2\pi = 5.3$ kHz and $10$ kHz, $Q_\text{m} \sim 5300$ and $5550$ near the Akhiezer limit, and record high frequencies $f_\text{m} = 8.5$ and $9.6$ GHz for microdisk optomechanical resonators. Further study is required to positively identify these modes with those from simulation, and to assess their $Q_\text{m}$ at cryogenic temperatures where thermal dissipation is reduced.

The modes measured here also present new opportunities for spin-mechanical coupling. The microscopic strain fields generated by phonons populating the microdisks's nanomechanical resonances can be coherently coupled to the electronic states of colour centers in diamond such as nitrogen vacancies \cite{ref:macquarrie2013msc, ref:ovartchaiyapong2014dsc, ref:teissier2014scn, ref:golter2016oqc, shandilya2021optomechanical} and silicon vacancies \cite{ref:maity2020cac}, allowing their spin states to be manipulated. The strength of spin-phonon coupling for a given mechanical mode is proportional to the peak stress per-phonon of displacement within the diamond crystal. We find that several of the modes measured here have more concentrated stress-per phonon than the radial breathing mode. Notably, as described in the Supplementary Information, the 6.1 GHz mode (simulated frequency) has stress components reaching $2.32$ kPa per phonon, which is larger than the peak value of $0.66$ kPa for the radial breathing mode. However, this mode is predicted to have a clamping loss limited $Q_\text{m} < 10^4$. The 8.15 GHz mechanical mode is predicted to have stress per phonon of 1.6 kPa, strong optomechanical coupling $g_0 \sim 25\,\text{kHz}$, and clamping loss limited $Q_\text{m} > 10^5$. Future work is required to positively identify this mode from the measured spectrum and to assess its potential for achieving high $Q_\text{m}$ in absence of thermal dissipation. 

The modes measured here will also enable a wider frequency range of resonant spin-phonon interactions compared with previous spin-optomechanics studies reliant upon the radial breathing mode \cite{shandilya2021optomechanical}. Their higher frequency will also be helpful for implementing non-resonant coupling schemes \cite{ref:neuman2021pi}. 

\section{Conclusion}

In summary, we have used modulation of intracavity radiation pressure to perform spectroscopy on a wide range of mechanical resonances of diamond microdisk cavities. The frequencies of these resonances extend to 10 GHz, far above that of the previously studied fundamental radial breathing mode of these devices. From the Fano nature of the measured optomechanical susceptibility, we extract both the optical Kerr nonlinearity of the diamond microdisk and the optomechanical coupling rate of the mechanical modes to the optical drive and probe modes. Using finite element method simulations and nanomechanical dissipation theory, we predict that at lower temperatures, the mechanical $Q$-factors of these high-frequency mechanical modes will be enhanced. The combination of large $g_0$, high frequency, and potential for high $Q_\text{m}$ will facilitate the simultaneous achievement of operating in the sideband resolved regime and with optomechanical cooperativity greater than unity, which are important conditions for quantum optomechanics experiments. This system has the potential to be used in a wide range of applications such as multimode optomechanics \cite{fiore2013optomechanical, lake2021processing}, spin-optomechanics \cite{shandilya2021optomechanical}, wavelength conversion \cite{ref:hill2012cow, mitchell2019optomechanically}, quantum phononics, and fundamental studies of diamond's nanomechanical properties. 

\begin{acknowledgements}
The authors gratefully acknowledge support from Alberta Innovates (Strategic Research Project) and NSERC (Discovery Grant). The authors thank David Lake and Matthew Mitchell for the fabrication of the device used in this work.
\end{acknowledgements}

\bibliography{nano_bib}

\end{document}